\documentclass[letterpaper, 10 pt, conference]{ieeeconf}  

\IEEEoverridecommandlockouts                              
\usepackage{graphicx,epstopdf}
\usepackage[cmex10]{amsmath}
\usepackage{amssymb,bbm}
\usepackage{algorithmic}
\usepackage{array}
\usepackage[caption=false,font=footnotesize]{subfig}
\usepackage{etoolbox}
\AtBeginEnvironment{bmatrix}{\setlength{\arraycolsep}{1pt}}
\title{\LARGE \bf Differential Dynamic Programming for Time-Delayed Systems}

\author{ \parbox{3 in}{\centering David D. Fan$^1$ and Evangelos A. Theodorou$^2$}
        \thanks{$^1$Institute for Robotics and Intelligent Machines, Georgia Institute of Technology, Atlanta, GA 30332-0150, USA. {\tt\small david.fan@gatech.edu}}
        \thanks{$^2$Daniel Guggenheim School of Aerospace Engineering, Georgia Institute of Technology, Atlanta, GA 30332-0150, USA. {\tt\small evangelos.theodorou@ae.gatech.edu}}
}

\DeclareMathOperator*{\argmin}{arg\,min}

\begin{document}

\maketitle
\thispagestyle{empty}
\pagestyle{empty}

\begin{abstract}
Trajectory optimization considers the problem of deciding how to control a dynamical system to move along a trajectory which minimizes some cost function.  Differential Dynamic Programming (DDP) is an optimal control method which utilizes a second-order approximation of the problem to find the control.  It is fast enough to allow real-time control and has been shown to work well for trajectory optimization in robotic systems. 
Here we extend classic DDP to systems with multiple time-delays in the state.  Being able to find optimal trajectories for time-delayed systems with DDP opens up the possibility to use richer models for system identification and control, including recurrent neural networks with multiple timesteps in the state.  We demonstrate the algorithm on a two-tank continuous stirred tank reactor.  We also demonstrate the algorithm on a recurrent neural network trained to model an inverted pendulum with position information only.
\end{abstract}

\section{INTRODUCTION}
Trajectory optimization, or more broadly known as optimal control, deals with the problem of finding a control input to a system which is optimal in some sense, i.e. with respect to a cost function.  Optimal control algorithms are often derived for systems with dynamics described by a first-order recurrence equation, where the next state is a function of the current state and control.  However, many real-world systems cannot be easily described this way, and may contain delays in controls or states.  For example, delays can be caused by communication delays in a distributed system, measurement delays from instrumentation, or from time-varying dynamics.  Some real-world examples include chemical processes, pneumatic systems with long transmission lines, hydraulic systems, and soft robotics with elastic members.  Furthermore, it may be easier to construct accurate dynamics models of real systems if some past history can be incorporated into the state, since the Markov assumption may often be too restrictive.  Finding algorithms for trajectory optimization on time-delay systems is an important problem, particularly with respect to robotics - where distributed, swarm, and non-rigid-body robotics, as well as modeling robots with recurrent neural networks are all quickly growing fields.

  Trajectory optimization is often performed on systems with initially unknown dynamics, i.e. the dynamics must be learned.  Creating models to approximate a real system's dynamics is a nontrivial problem.  Often in optimal control, the dynamics are assumed to be known, and fully described by a first-order differential equation.  In practice, however, it may be difficult to create an reliable model of unknown dynamics using only a first-order differential model.  Parametric models which predict the next state given a short history of states may be more flexible and powerful.  Examples include NARX models, time-delay neural networks, or other customized neural network architectures.  In Lenz et al.'s DeepMPC work, a deep neural network was used to model the dynamics of a robot cutting various foods.  Their network used a time-block design, in which time-series data was partitioned into blocks and fed into the neural network\cite{Lenz}.  In work by Whalstr\"om et al., a dynamical model was learned to map a sequence of images to control torques, with the neural network taking several past images as input \cite{Wahlstrom2015a}.  Finally, a time-delay neural network was used to model helicopter blade torsions and flapping in \cite{Belo}.

  Broadly speaking, there are two well-studied approaches to trajectory optimization - one is that of dynamic programming, and the other relies on Pontryagin's maximum principle.  For systems with time delay, much work has been done using Pontryagin's maximum principle, beginning with Kharatishvili in 1961 \cite{GL1961}.  A maximum principle approach by Guinn first reduces the delayed problem to a non-delayed one, then proceeds normally \cite{Guinn1976}.  More recently, G\"ollmann et al. provided necessary optimality conditions for delayed problems with control and state constraints, as well as giving a review of maximum principles for time-delay systems \cite{Gollmann2009}.
  
  As for utilizing dynamic programming for time-delayed systems, some work has been done which focused on Iterative Dynamic Programming (IDP) \cite{Dadebo1992}\cite{Hwang1999}.  IDP utilizes a grid-search in the controls and iteratively shrinks the grid size and space in an effort to find the globally optimal control.  This approach is severely affected by the curse of dimensionality, and will not work for problems with a larger number of states.  A dynamic programming method which avoids the need to discretize the state and control space is Differential Dynamic Programming (DDP), which iteratively solves a second-order local approximation of the problem \cite{Mayne1966}.  It has the advantage of providing both feed-forward and feedback control gains, and is fast enough to support real-time control of a humanoid robot \cite{Tassa}.  A simpler variant of DDP, called the iterative Linear Quadratic Gaussian (iLQG) method, considers only first-order dynamics, and is well studied \cite{Todorov2005}.  Several extensions to DDP have been recently made, including control-limited DDP \cite{Tassa2014}, stochastic DDP \cite{Theodorou2010}, combining optimal control and optimal estimation \cite{Kobilarov}, and probabilistic DDP, which controls the belief space for systems with unknown dynamics \cite{Pan2014a}.  DDP has also been demonstrated to work well for receding horizon control in robotic systems \cite{Tassa2008a}.
  
In this work we extend Differential Dynamic Programming to systems with multiple delays in the state.  This opens up a range of possibilities in terms of performing trajectory optimization on time-delayed dynamical systems.  Of special interest is the case where a system's dynamics are unknown but can be approximated with a parametric model containing state delays, e.g. a neural network.

   An overview of the paper is as follows: Section II presents the derivation of Delayed DDP.  Section III describes the algorithm based on the derivation.  Section IV discusses implementation of the algorithm on a two-stage continuously stirred tank reactor system.  Section V discusses modeling dynamical systems with delayed recurrent neural networks, and demonstrates Delayed DDP on a learned neural network model for an inverted pendulum model using position information.

\section{DELAYED DIFFERENTIAL DYNAMIC PROGRAMMING}
\subsection{Problem Formulation}
Let the sequence \(\{x_i\}\) be a state trajectory comprised of states \(x_i\in \mathbb{R}^n\) for times \(i=0,\dotsc,N\).  The trajectory is determined by the k-th order difference equation:
\begin{multline}
\label{eq:fdelay}
x_{i+1}=f(x_i,x_{i-1},\dotsc,x_{i-k},u_i), \quad i=0,\dotsc,N-1\\
x_{-j}=x^0_{-j}, \quad j=0,\dotsc,k, \quad 0<k\ll N
\end{multline}
where \(\{u_i\}\) is a control sequence with \(u_i\in \mathbb{R}^m\) for times \(i=0,\dotsc,N-1\), \(f\) maps \(\mathbb{R}^{nk}\times\mathbb{R}^m\mapsto\mathbb{R}^n\) and is twice differentiable, and \(x^0_{-j}\) are the initial delayed states.  The dynamics depend on the past \(k+1\) states as well as the controls at the current time.  Let \(\mathbf{\bar{x}}_i\) denote the sequence of states \(\{x_i,x_{i-1},\dotsc,x_{i-k}\}\).  The initial condition is given by \(\mathbf{\bar{x}}_0^0\).  Then we can write (\ref{eq:fdelay}) as:
\begin{align}
x_{i+1}&=f(\mathbf{\bar{x}}_i,u_i), \quad i=0,\dotsc,N-1\\
\mathbf{\bar{x}}_0&=\mathbf{\bar{x}}_0^0\nonumber
\end{align}
Define a cost function as
\begin{equation}
J^0(\mathbf{\bar{x}}_0,\mathcal{U})=\sum_{j=0}^{N-1}L^j(\mathbf{\bar{x}}_j,u_j)+L^N(\mathbf{\bar{x}}_N)
\end{equation}
where \(\mathcal{U}=\{u_i\}, i=0,\dotsc,N-1\), and \(L^j\) are twice-differentiable nonnegative scalar functions.  The problem of optimal control is to find a \(\mathcal{U}\) that minimizes this cost function:
\begin{equation}
\mathcal{U}^{*}=\argmin_{\mathcal{U}}J^0(\mathbf{\bar{x}}_0,\mathcal{U})
\end{equation}

\subsection{Bellman Equation with Delays}
The Bellman equation is a necessary condition of optimality for the dynamic programming problem.  In classic DDP without delays, the Taylor expansion of the Bellman equation about the point \((x_i,u_i)\) is taken at each timestep.  For the case with delays, the Taylor expansion must be taken around the segment of past history within the delay, i.e., about \(\mathbf{\bar{x}}_i=(x_i,x_{i-1},\dotsc,x_{i-k},u_i)\).  This follows the idea of Guinn whereby a delayed system is reduced to one without delays \cite{Guinn1976}.\\
Define the cost-to-go function as
\begin{equation}
\label{eq:costtogodelay}
J^i(\mathbf{\bar{x}}_i,\mathcal{U}_i)=\sum_{j=i}^{N-1}L^j(\mathbf{\bar{x}}_j,u_j)+L^N(\mathbf{\bar{x}}_N)
\end{equation}
Minimizing (\ref{eq:costtogodelay}) with respect to the current and all future controls \(\mathcal{U}_i\) gives an expression for the value function:
\begin{equation}
V^i(\mathbf{\bar{x}}_i)=\min_{\mathcal{U}_i}\sum_{j=i}^{N-1}L^j(\mathbf{\bar{x}}_j,u_j)+L^N(\mathbf{\bar{x}}_N)
\end{equation}
This expression can be written iteratively, yielding a Bellman equation for delayed systems:
\begin{equation}
\label{eq:bellmandelay}
V^i(\mathbf{\bar{x}}_i)=\min_{u_i}[L^i(\mathbf{\bar{x}}_i,u_i)+V^{i+1}(\mathbf{\bar{x}}_{i+1})]
\end{equation}
Note that \(\mathbf{\bar{x}}_{i+1}\) is simply \(\{f(\mathbf{\bar{x}}_i,u_i),x_i,x_{i-1},\dotsc,x_{i-k+1}\}\).  So the rightmost term of (\ref{eq:bellmandelay}) is a function of \(\mathbf{\bar{x}}_i\) still.  Now, as in the case of classic DDP without delays, we can approximate the argument of the minimum in (\ref{eq:bellmandelay}) via a second-order Taylor expansion to find \(u\).

\subsection{Quadratic Approximation}
Define \(Q\) as the argument of the minimum of (\ref{eq:bellmandelay}) (where dependence on time \(i\) is implicit), and write it as a function of perturbations around \((x_i,x_{i-1},\dotsc,x_{i-k},u_i)\).  Expanding via the second order Taylor expansion gives
\begin{multline}
\label{eq:taylordelay}
Q(\mathbf{\bar{x}}_i+\delta\mathbf{\bar{x}}_i,u_i+\delta u_i)-Q(\mathbf{\bar{x}}_i,u_i)\\
\approx\sum_{j=0}^kQ_{x_{i-j}}\delta x_{i-j}+Q_{u_i}\delta u_i +\\
\frac{1}{2}
\begin{bmatrix}
\delta x_i\\
\vdots\\
\delta x_{i-k}\\
\delta u_i
\end{bmatrix}^\intercal
\begin{bmatrix}
Q_{x_ix_i} & \cdots & Q_{x_ix_{i-k}} & Q_{x_iu_i}\\
\vdots & \ddots & \vdots & \vdots\\
Q_{x_{i-k}x_i} & \cdots & Q_{x_{i-k}x_{i-k}} & Q_{x_{i-k}u_i}\\
Q_{u_ix_i} & \cdots & Q_{u_ix_{i-k}} & Q_{u_iu_i}
\end{bmatrix}
\begin{bmatrix}
\delta x_i\\
\vdots\\
\delta x_{i-k}\\
\delta u_i
\end{bmatrix}
\end{multline}
where the subscripts on \(Q\) indicate partial derivatives.  To find these coefficients of the Taylor expansion we must expand both \(L^i(\mathbf{\bar{x}}_i,u_i)\) and \(V^{i+1}(\mathbf{\bar{x}}_{i+1})\) in (\ref{eq:bellmandelay}).  Expanding \(L^i(\mathbf{\bar{x}}_i,u_i)\), we have:
\begin{multline}
\label{eq:quadratic0}
L^i(\mathbf{\bar{x}}_i+\delta\mathbf{\bar{x}}_i,u_i+\delta u_i)-L^i(\mathbf{\bar{x}}_i,u_i)\\
\approx L^i_{x_i}\delta x_i+L^i_{x_{i-1}}\delta x_{i-1}+\dotsc+L^i_{x_{i-k}}\delta x_{i-k}+L^i_{u_i}\delta u_i+\\
\frac{1}{2}
\begin{bmatrix}
\delta x_i\\
\vdots\\
\delta x_{i-k}\\
\delta u_i
\end{bmatrix}^\intercal
\begin{bmatrix}
L^i_{x_ix_i} & \cdots & L^i_{x_ix_{i-k}} & L^i_{x_iu_i}\\
\vdots & \ddots & \vdots & \vdots\\
L^i_{x_{i-k}x_i} & \cdots & L^i_{x_{i-k}x_{i-k}} & L^i_{x_{i-k}u_i}\\
L^i_{u_ix_i} & \cdots & L^i_{u_ix_{i-k}} & L^i_{u_iu_i}
\end{bmatrix}
\begin{bmatrix}
\delta x_i\\
\vdots\\
\delta x_{i-k}\\
\delta u_i
\end{bmatrix}
\end{multline}
The expansion of \(V^{i+1}(\mathbf{\bar{x}}_{i+1})\) requires more careful consideration.  We will use the following notation to allow us to drop the indices \(i\):  The partial derivative \(V^{i+1}_{x_{i+1}}\) is the derivative of the value function at time \(i+1\) with respect to the first argument, \(x_{i+1}\).  We can write this in short hand as \(V'_0\), where the \('\) denotes the value function at time \(i+1\).  Similarly, write \(V^{i+1}_{x_{i}}\) as \(V'_1\), up to \(V^{i+1}_{x_{i-k+1}}\) as \(V'_k\).  For second derivatives, write \(V^{i+1}_{x_{i+1},x_{i+1}}\) as \(V'_{00}\), etc.  The expansion yields:
\begin{multline}
\label{eq:quadratic1}
V^{i+1}(\mathbf{\bar{x}}_{i+1}+\delta\mathbf{\bar{x}}_{i+1})-V^{i+1}(\mathbf{\bar{x}}_{i+1})\\
\approx V'_0\delta x_{i+1}+V'_1\delta x_{i}+\dotsc+V'_k\delta x_{i-k+1}\\
+\frac{1}{2}
\begin{bmatrix}
\delta x_{i+1}\\
\vdots\\
\delta x_{i-k+1}
\end{bmatrix}^\intercal
\begin{bmatrix}
V'_{00} & \cdots & V'_{0k}\\
\vdots & \ddots & \vdots\\
V'_{k0} & \cdots & V'_{kk}
\end{bmatrix}
\begin{bmatrix}
\delta x_{i+1}\\
\vdots\\
\delta x_{i-k+1}
\end{bmatrix}
\end{multline}
We can find the expression for \(\delta x_{i+1}\) by expanding the function \(f\) to the second order as well:
\begin{multline}
\label{eq:quadratic2}
\delta x_{i+1}=f(\mathbf{\bar{x}}_i+\delta\mathbf{\bar{x}}_i,u_i+\delta u_i)-f(\mathbf{\bar{x}}_i,u_i)\\
\approx f_{x_i}\delta x_i+f_{x_{i-1}}\delta x_{i-1}+\dotsc+f_{x_{i-k}}\delta x_{i-k}+f_{u_i}\delta u_i\\
+\frac{1}{2}
\begin{bmatrix}
\delta x_i\\
\vdots\\
\delta x_{i-k}\\
\delta u_i
\end{bmatrix}^\intercal
\begin{bmatrix}
f_{x_ix_i} & \cdots & f_{x_ix_{i-k}} & f_{x_iu_i}\\
\vdots & \ddots & \vdots & \vdots\\
f_{x_{i-k}x_i} & \cdots & f_{x_{i-k}x_{i-k}} & f_{x_{i-k}u_i}\\
f_{u_ix_i} & \cdots & f_{u_ix_{i-k}} & f_{u_iu_i}
\end{bmatrix}
\begin{bmatrix}
\delta x_i\\
\vdots\\
\delta x_{i-k}\\
\delta u_i
\end{bmatrix}
\end{multline}
Plugging (\ref{eq:quadratic2}) into (\ref{eq:quadratic1}) gives a summation of terms which are from first to fourth order with respect to \(\delta\mathbf{\bar{x}}_i=(\delta x_i,\delta x_{i-1},\dotsc,\delta x_{i-k},\delta u_i)\).  Since we are only interested in a second order approximation, we can drop the third and forth order terms, leaving a summation of terms which are either first or second order.  The expressions for the coefficients \(Q\) in (\ref{eq:taylordelay}) are found after collecting these terms multiplied by the same \(\delta x_{i-j}\), along with the terms in (\ref{eq:quadratic0}).  Considering the first order coefficients, for \(j=0,\dotsc,k-1\), we have:
\begin{equation}
Q_{x_{i-j}}=L^i_{x_{i-j}}+f^\intercal_{x_{i-j}}V'_0+V'_{j+1}
\end{equation}
When \(j=k\), since (\ref{eq:quadratic1}) has no \(\delta x_{i-k}\) term, the \(V'_{j+1}\) disappears:
\begin{equation}
Q_{x_{i-k}}=L^i_{x_{i-k}}+f^\intercal_{x_{i-k}}V'_0
\end{equation}
Also, gathering up the terms corresponding to \(\delta u_i\) gives us
\begin{equation}
Q_{u_i}=L^i_{u_i}+f^\intercal_{u_i}V'_0
\end{equation}
To simplify notation further, write \(f_{x_{i-j}}\) as \(f_{x^j}\), and follow the same pattern for \(L^i_{x^j}\) and \(Q_{x^j}\).  The full expressions for the Taylor coefficients \(Q\) for both first and second order are, for \(j,l=0,\dotsc,k\):
\begin{subequations}
\label{eq:coeffdelay}
\begin{align}
Q_{x^j}&=L_{x^j}+f^\intercal_{x^j}V'_0 + \mathbf{1}_{j\neq k}V'_{j+1}\\
Q_{u}&=L_{u}+f^\intercal_{u}V'_0\\
Q_{x^ju}&=L_{x^j,u}+f^\intercal_{x^j}V'_{0,0}f_u+V'_0\cdot f_{x^ju}+\mathbf{1}_{j\neq k}V'_{j+1,0}f_u\\
Q_{uu}&=L_{u,u}+f^\intercal_{u}V'_{0,0}f_{u}+V'_0\cdot f_{u,u}\\
Q_{x^jx^l}&=L_{x^j,x^l}+f^\intercal_{x^j}V'_{0,0}f_{x^l}+V'_0\cdot f_{x^jx^l}\\
&+\mathbf{1}_{j\neq k}V'_{j+1,0}f_{x^l}+\mathbf{1}_{l\neq k}f_{x^j}^\intercal V'_{0,l+1}+\mathbf{1}_{j,l\neq k}V'_{j+1,l+1}\nonumber
\end{align}
\end{subequations}
where the dot \(\cdot\) denotes contraction with a tensor and \(\mathbf{1}_{j\neq k}\) is the indicator function, taking a value of \(1\) when \(j\neq k\) and \(0\) otherwise.  The tensor contractions arise since \(f\) is a vector-valued function, so its first derivative is a matrix, and its second derivative is a tensor of rank 3.  We can write the tensor contraction explicitly as the sum of each element of the vector \(V'_0\) times the Hessian of the corresponding element of \(f\):
\begin{equation} 
V'_0\cdot f_{x^jx^l}=\sum^n_{p=1} {V'_0}^{(p)}f^{(p)}_{x^jx^l}
\end{equation}

\subsection{Backward and Forward Pass}
Now we can find an expression for the locally optimal control.  Minimizing (\ref{eq:taylordelay}) with respect to \(\delta u\) gives
\begin{equation}
\label{eq:policydelay1}
\delta u^{*}=-Q_{uu}^{-1}\Big(Q_u+\sum_{j=0}^kQ_{x^ju}^\intercal\delta x^j\Big)
\end{equation}
This is true as long as \(Q_{uu}\) is positive-definite.  If \(Q_{uu}\) is not positive definite, the standard regularization proposed in \cite{Mayne1966} is to add a diagonal term to the Hessian:
\begin{equation}
\label{eq:policydelay2}
\tilde{Q}_{uu}=Q_{uu}+\mu\mathbf{I}
\end{equation}
So we have a linear control law with feedback gains which depends on the past \(k\) timesteps.\\
Define the open-loop gain \(\mathbf{k}=-\tilde{Q}_{uu}^{-1}Q_u\) and the feedback gains \(\mathbf{K}_j=-\tilde{Q}_{uu}^{-1}Q_{x^ju}^\intercal\) for \(j=0,\dotsc,k\).  Plugging the control policy (\ref{eq:policydelay1}) into (\ref{eq:taylordelay}) and (\ref{eq:bellmandelay}) and gathering terms which are zeroth, first, and second order in $\delta x$ gives recursive expressions for the quadratic approximation of the value at each timestep.  Doing so yields, for \(j,l=0,\dotsc,k\):
\begin{subequations}
\label{eq:valueupdatedelay}
\begin{align}
\Delta V&=\frac{1}{2}\mathbf{k}^\intercal \tilde{Q}_{uu}\mathbf{k}+\mathbf{k}^\intercal Q_u\\
V_{j}&=Q_{x^j}+\mathbf{K}_j^\intercal \tilde{Q}_{uu}\mathbf{k}+\mathbf{K}_j^\intercal Q_u+Q_{x^ju}\mathbf{k}\\
V_{j,l}&=Q_{x^jx^l}+Q_{x^ju}\mathbf{K}_l+\mathbf{K}_j^\intercal Q_{ux^l}+\mathbf{K}_j^\intercal \tilde{Q}_{uu}\mathbf{K}_l
\end{align}
\end{subequations}
Now in the same manner as classic DDP, one may first compute a forward pass with an nominal control trajectory, then compute a backwards pass of the values of \(V\), \(V_j\), and \(V_{j,l}\) along with the control policy \(\{\mathbf{k},\mathbf{K}_j\}\) at each timestep, starting at the last timestep and working backwards.  When starting with the last timestep, the boundary conditions for V are given by the partial derivatives of \(L^N(\mathbf{\bar{x}}_N)\):
\begin{subequations}
\begin{align}
V_j^N&=L^N_{x^j}\\
V_{j,l}^N&=L^N_{x^j,x^l}
\end{align}
\end{subequations}
The forward pass is then calculated as:
\begin{subequations}
\label{eq:forwardpassdelay}
\begin{align}
\mathbf{\hat{\bar{x}}}_0&=\mathbf{\bar{x}}^0_0\\
\hat{u}_i&=u_i+\alpha\mathbf{k}(i)+\sum_{j=0}^k\mathbf{K}_j(i)(\hat{x}_{i-j}-x_{i-j})\\
\hat{x}_{i+1}&=f(\mathbf{\hat{\bar{x}}}_i,\hat{u}_i)
\end{align}
\end{subequations}
where \(0<\alpha\leq1\) is used to keep the new trajectory close to the old one, since the quadratic approximation is only valid near the nominal trajectory.

\section{ALGORITHM}
Implementation follows the classic DDP case which is covered by Tassa et al. \cite{Tassa} in detail.  The line search parameter \(0<\alpha\leq1\) is used to scale the open-loop gain $\mathbf{k}$.  A quadratic schedule is used to change the regularization parameter \(\mu\) to find a good value if needed.  A single iteration consists of three steps:
\begin{enumerate}
\item \textbf{Derivatives:} Compute derivatives of \(L\) and \(f\) in (\ref{eq:coeffdelay}).
\item\textbf{Backward Pass:} For \(i=N-1,\dotsc,0\), iteratively calculate the new control policy from (\ref{eq:coeffdelay},\ref{eq:valueupdatedelay}).  If for any $i$ a non-positive-definite \(\tilde{Q}_{uu}\) is found, increase \(\mu\) and restart the backwards pass.  Otherwise, decrease \(\mu\).
\item\textbf{Forward Pass:} Set \(\alpha=1\), then iterate (\ref{eq:forwardpassdelay}) forward for $i=0,\dotsc,N-1$.  If the integration diverges or if the actual cost reduction is less than expected (see \cite{Tassa}), reduce \(\alpha\) and restart the forward pass.
\end{enumerate}
The computational cost increase of Delayed DDP compared to classic DDP is proportional to the size of the delay.  Assuming \(n=m\), the computational cost of one iteration of classic DDP is \(\mathcal{O}(Nn^3)\).  The cubic dependence on the dimension of the states and control come from multiplying matrices when back-propagating the value function, as well as the matrix inversion of \(Q_{uu}\).  With Delayed DDP, the computational cost is increased to \(\mathcal{O}(k^2Nn^3)\).  Therefore, a short delay is preferable to a long one.

Classic DDP converges quadratically to a local minimum of the cost function \cite{Mayne1966}.  Delayed DDP also converges quadratically; this can be seen because the delayed problem is collapsed to a non-delayed problem in (\ref{eq:bellmandelay}), and the convergence analysis that follows is the same as in that of classic DDP.  If fast computation speed is desired over quadratic convergence, the second-order dynamics terms containing tensor contractions can be dropped, resulting in an iLQG algorithm for time-delay systems.  This may be advantageous for the receding horizon case, as in Model-Predictive Control (MPC), where less-than-optimal solutions are more acceptable and having low computation time takes a higher priority \cite{Tassa2008a}.

\section{EXPERIMENTS}
\subsection{Delayed DDP Applied to Two-Stage Continuously Stirred Tank Reactor System}
\begin{figure}
\centering
\includegraphics[width = 0.48 \textwidth]{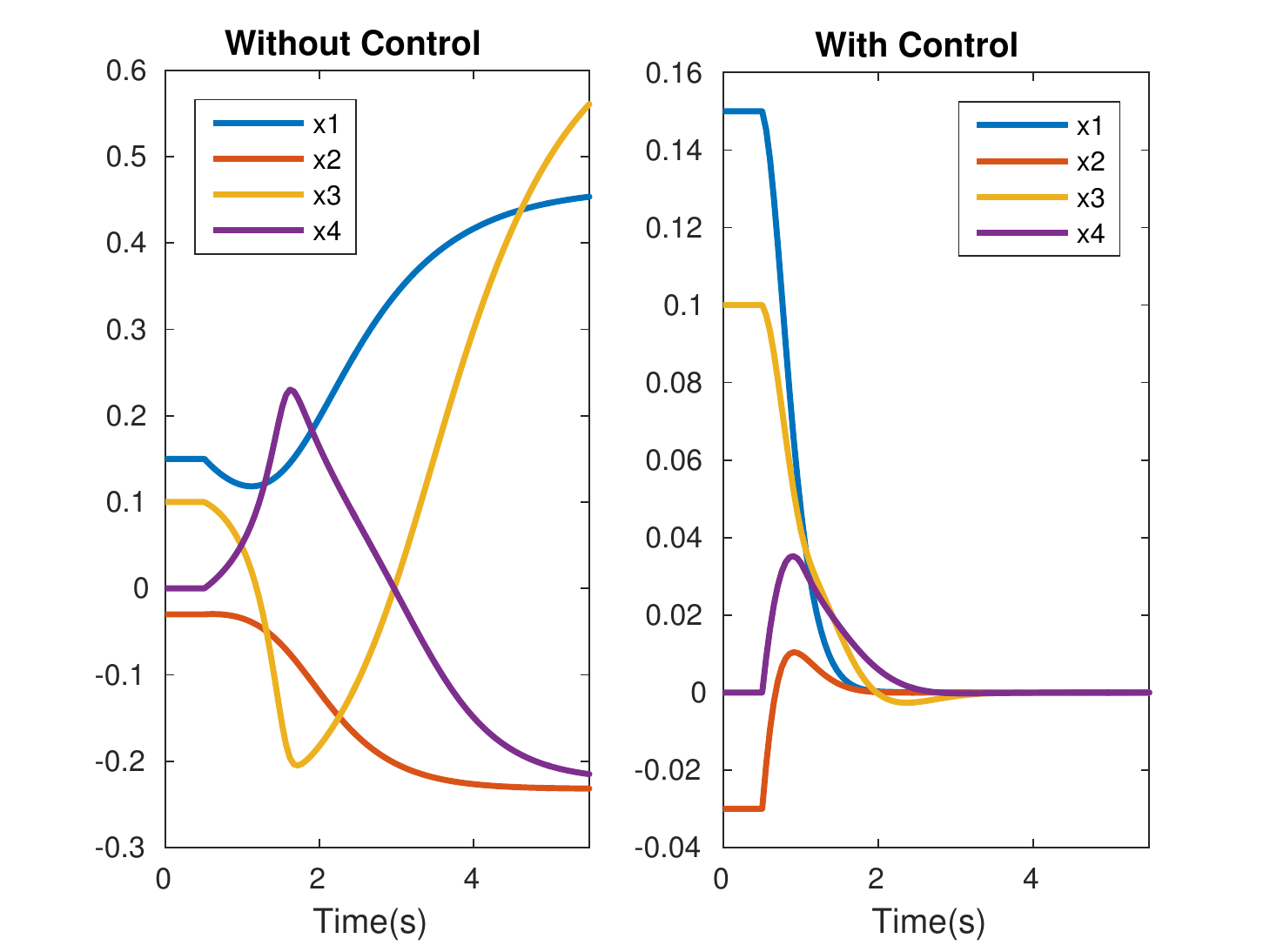}
\caption{Results for Delay DDP on a two-stage continuously stirred tank reactor system with a state delay of \(0.5\) seconds.  Left: State trajectories without any control.  Right: State trajectories with control found by Delayed DDP algorithm.  All states are successfully pushed to \(0\).}
\label{twotank1}
\end{figure}

\begin{figure}
\centering
\includegraphics[width = 0.48 \textwidth]{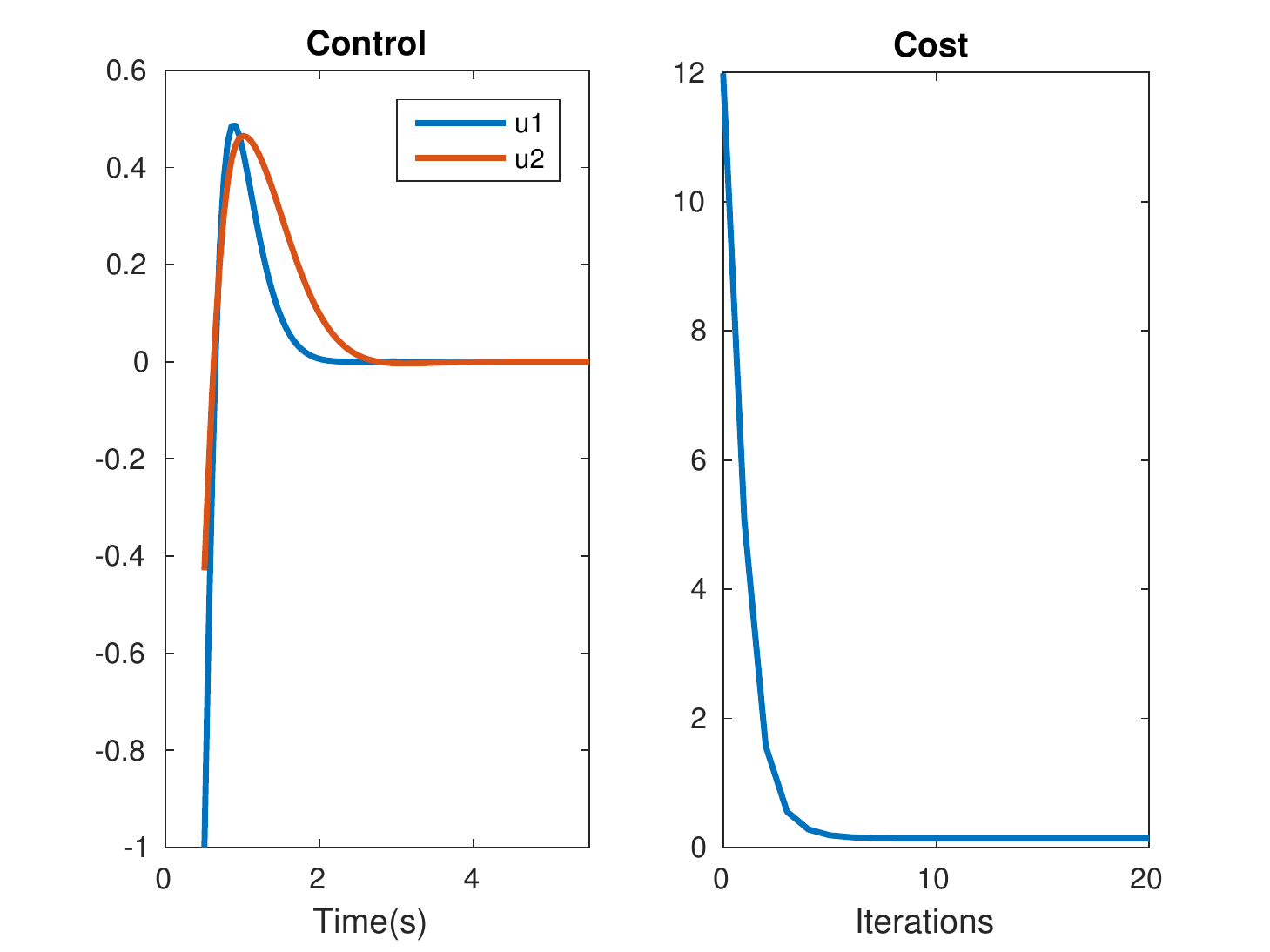}
\caption{Results for Delay DDP on a two-stage continuously stirred tank reactor system with a state delay of \(0.5\) seconds.  Left: Optimal control sequence.  Right: Total cost per iteration of Delayed DDP algorithm.}
\label{twotank2}
\end{figure}

We simulate the Delayed DPP algorithm on a nonlinear two-stage continuously stirred tank reactor system (CSTR) \cite{Dadebo1992}.  The system has four states and is under-actuated with two control inputs.  The 1st and 3rd states correspond to normalized concentrations, and the 2nd and 4th states correspond to normalized temperatures.  The system is described by:
\begin{align*}
\frac{dx_1(t)}{dt}&=0.5-x_1(t)-R_1\\
\frac{dx_2(t)}{dt}&=-2(x_2(t)+0.25)-u_1(t)(x_2(t)+0.25)+R_1\\
\frac{dx_3(t)}{dt}&=x_1(t-\tau)-x_3(t)-R_2+0.25
\end{align*}
\begin{equation}
\frac{dx_4(t)}{dt}=x_2(t-\tau)-2x_4(t)-u_2(t)(x_4(t)+0.25)+R_2-0.25
\end{equation}
with
\begin{align*}
R_1&=(x_1(t)+0.5)\exp\Big(\frac{25x_2(t)}{x_2(t)+2}\Big)\\
R_2&=(x_3(t)+0.25)\exp\Big(\frac{25x_4(t)}{x_4(t)+2}\Big)
\end{align*}
and an initial state
\begin{equation*}
\mathbf{\bar{x}}^0(t)=[0.15, -0.03, 0.1, 0.0]^\intercal, \quad t\in[-\tau,0]
\end{equation*}
where \(\tau\) is the time delay and was set to \(0.5\) seconds.  Euler's discretization was used with a step size of \(0.05\) seconds to bring the system into discrete equations, and a horizon of \(100\) timesteps, giving a \(5\) second horizon, was used.  We minimized a quadratic cost function in both states and controls:
\begin{equation}
L(\mathbf{\bar{x}}_i,u_i)=\frac{1}{2}\sum_{j=0}^k x_i^{j\intercal} P x_i^j+\frac{1}{2}u_i^\intercal R u_i
\end{equation}
where P and R are diagonal and positive semidefinite matrices.  We found that using a cost function which depends on the entire delay history rather than the current state alone improved the ease of finding a good solution.  The control cost was $R=0.1*I_{2x2}$ and state cost was $P=I_{4x4}$.  For simplicity, we used a fixed learning rate \(\alpha=0.4\) with no regularization, and omitted the second-order dynamics terms.  The algorithm converged to an acceptable solution after about \(5\) iterations; we show the results after \(20\) iterations (Figures \ref{twotank1} and \ref{twotank2}).  Including the second-order dynamics terms results in the same solution if some regularization is used and more iterations are taken.

\begin{figure}
\centering
\includegraphics[width = 0.48 \textwidth]{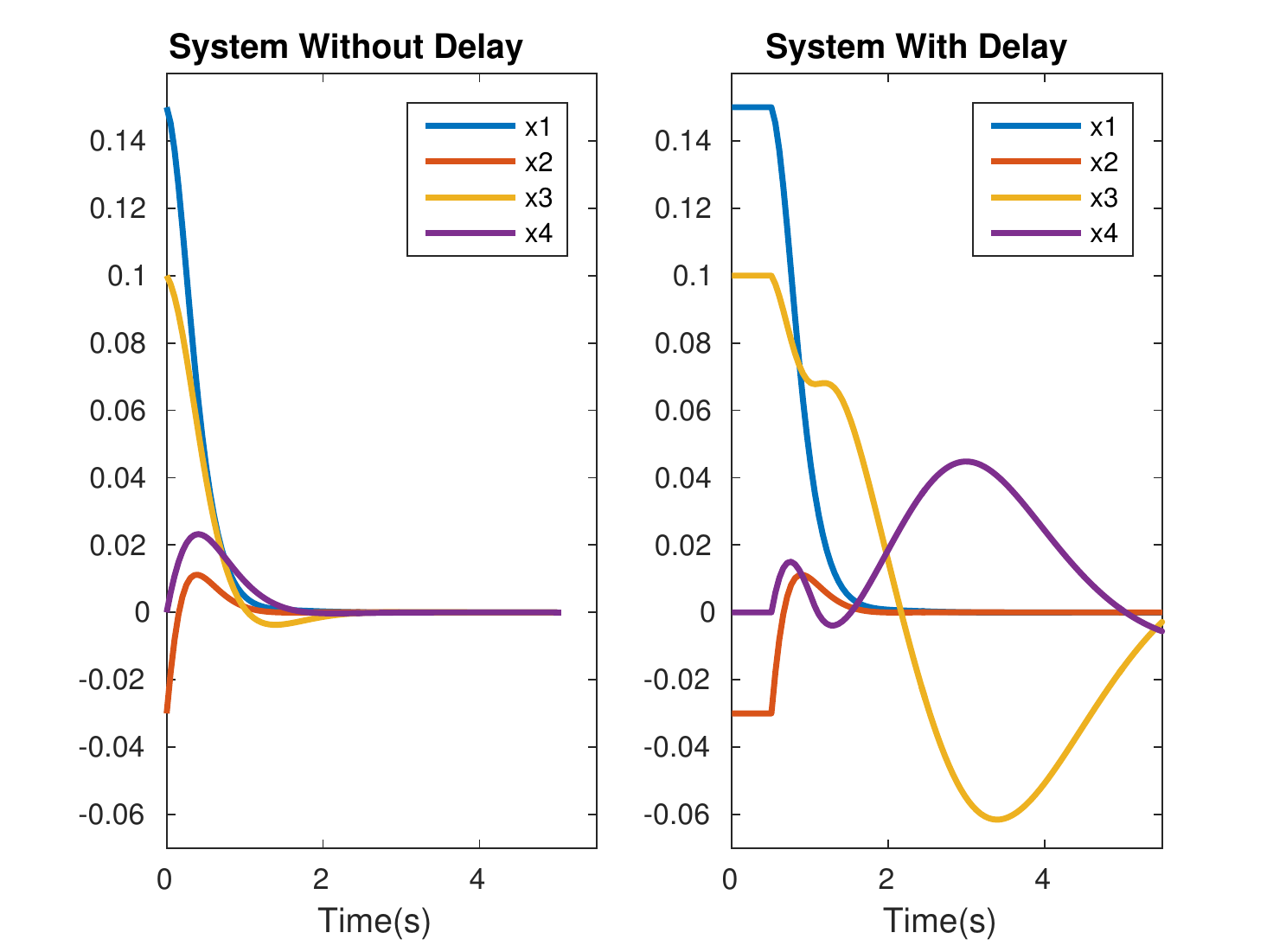}
\caption{Left: Classic DDP controls the system with no delays successfully.  Right: The same classic DDP control applied to a system with delays ($\tau=0.5$ seconds).  The classic DDP optimal control does not adequately control the system with delays.}
\label{delaycompare}
\end{figure}

\begin{figure}
\centering
\includegraphics[width = 0.48 \textwidth]{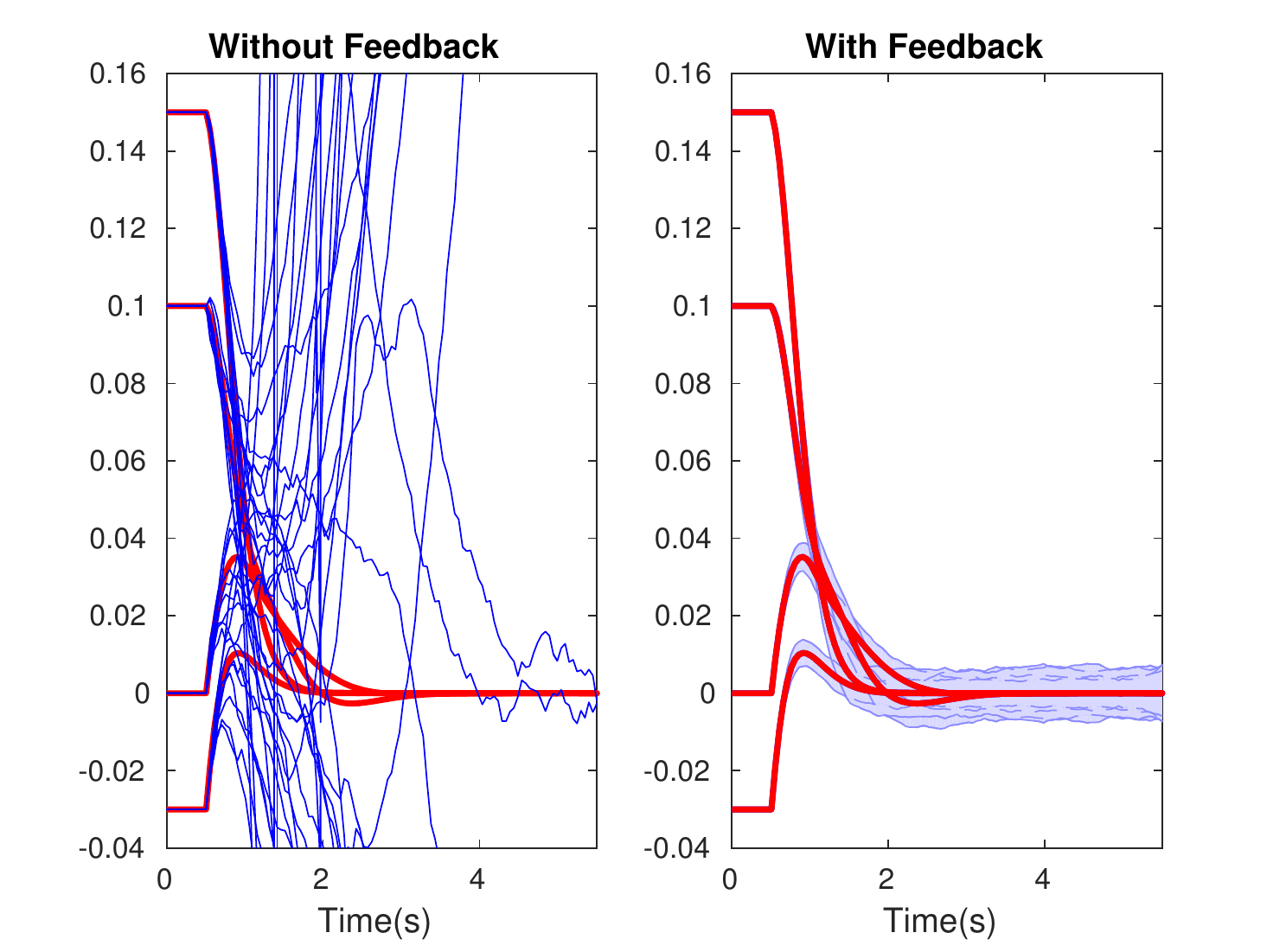}
\caption{Demonstration of optimal control with feedback gains.  Independent noise is added to each state with variance \(\sigma=0.01\).  Left: State trajectories without using feedback gains.  Red indicates trajectories without noise, blue indicate 10 sampled trajectories.  Right: State trajectories using feedback gains.  Blue error bars indicate standard error estimated from 100 sampled trajectories.}
\label{delaynoise}
\end{figure}

   To test the difference between Delayed DDP and classic DDP, we used classic DDP to find an optimal control for the two-stage CSTR system when \(\tau=0\), i.e. for the system without delays.  We then applied this optimal control an identical two-stage CSTR system except for having a delay of \(\tau=0.5\).  The classic DDP optimal control sequence was unable to adequately control the system with delays (Figure \ref{delaycompare}).  Therefore, using classic DDP on systems where delays play a role may be an inadequate approach.\\
A distinct advantage of DDP over other trajectory optimization techniques such as those relying on the maximum principle is that the DDP algorithm gives feedback gains \(\mathbf{K}_j\), which arise naturally from the back-propagation of the value function.  These feedback gains can be used to steer the system in the presence of noise or disturbances.  To demonstrate the value of having these feedback gains, we added some Wiener process noise to the two-stage CSTR system and ran it with and without feedback gains.  Independent noise was added to each state at each timestep drawn from the normal distribution $\mathcal{N}(0,\sigma\sqrt{dt})$.  Without feedback gains, the noisy control quickly steers the state trajectories off course, and the unstable dynamical system causes the states to explode (Figure \ref{delaynoise}).  However, using the feedback gains results in reliable control, keeping the states close to their intended trajectories.\\

\section{MODELING WITH DELAYED RECURRENT NEURAL NETWORKS}
Recurrent neural networks have been shown to be useful for approximating dynamical systems \cite{Polycarpou1991}\cite{Funahashi1993}\cite{Sutskever2010}.  Various architectures have been considered, including Temporal-Kernel Recurrent Neural Networks, Echo-State Neural Networks, Long-Short Term Neural Networks, and more (for a review, see \cite{Schmidhuber2014}).  More recently, some work has been done in building deep recurrent neural networks models to approximate the dynamics of various tasks such as cutting fruit with a robot arm \cite{Lenz}, controlling a pendulum from images \cite{Wahlstrom2015a}, or learning inverse dynamics of a musculoskeletal robot \cite{Hartmann2012}.  For each of these works, the authors found that in order to achieve good performance, it was necessary to train the neural networks on segments of delayed data, creating recurrent neural networks which are deep in time \cite{Pascanu2013}.  The necessity of making use of delays is especially evident for tasks such as controlling a system from images alone, since a single frame is insufficient for providing information about state velocities.  However, control and trajectory optimization of such delayed recurrent neural networks has not been previously addressed.  In the aforementioned works, the authors used Model Predictive Control along with policy gradient methods to find a control to accomplish some task.  This approach is computationally expensive and demands constantly re-querying the system and updating the trajectory.  Instead, our approach here is to use the Delayed DDP framework to efficiently plan an entire trajectory at once.  Feedback gains are naturally obtained from the back-propagation of the value function, which can be used to compensate for both modeling errors and control noise.  This approach should be both more computationally efficient and more powerful.
\begin{figure}
\centering
\includegraphics[width = 0.35 \textwidth]{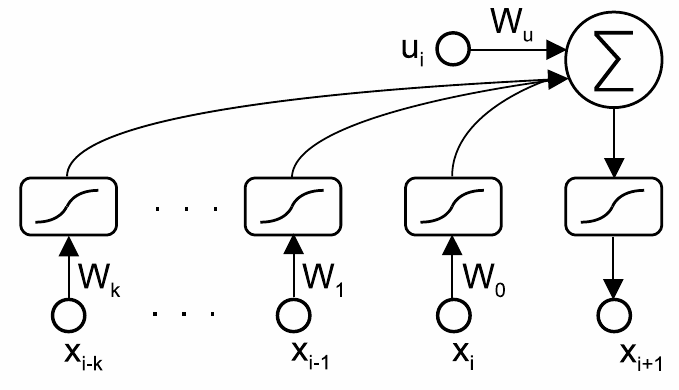}
\caption{Neural network architecture used to model system dynamics.  The next state \(x_{i+1}\) is a function of a history of past states and the current control input \(u_i\).}
\label{fig:nndiag}
\end{figure}
We use the following neural network architecture to approximate a dynamical system:
\begin{multline}
\label{eq:nn}
x_{i+1}=\sigma\bigg(W_uu_i+b_u+\sigma(W_0x_i+b_0)\\
+\sigma(W_1x_{i-1}+b_1)+\dotsc+\sigma(W_kx_{i-k}+b_k)\bigg)
\end{multline}
where \(\sigma\) is the activation function of choice (we use the hyperbolic tangent), and \(\{W_u,W_j\}\),\(\{b_u,b_j\}\) are weight matrices and bias vectors, respectively (Figure \ref{fig:nndiag}).  The network is trained to produce the state vector at the next timestep given the past \(k+1\) states and the current control input.  However, directly training this feed-forward architecture to do one-step prediction is unlikely to create a system which closely approximates the real system's dynamics.  This is because we are interested not only in one-step prediction but in multiple-step sequence prediction.  This problem of training a recurrent neural network to do multiple-step prediction has been previously addressed in various ways.  One approach, called Scheduled Sampling, is to gradually ease the training of the network on data alone to training the network on its own outputs \cite{Bengio2015a}.  Another more data-driven approach known as Data as Demonstrator augments the training data with the model's own errors \cite{Venkatraman2015}.  Here, we take a simpler approach.  The step forward function in (\ref{eq:nn}) is a feed-forward neural network, but since it takes inputs of past states, it can be considered a recurrent neural network.  We therefore train the network with an entire sequence of data, back-propagating the error function backwards through time, just as one would train any normal recurrent neural network.  We can also augment the state vector with hidden states to increase the expressiveness of the network.  Let \(\hat{x}_i\) be the augmented state, with \(\hat{x}_i=[x_i,h_i]\), \(x_i\in\mathcal{R}^n\), \(h_i\in\mathcal{R}^r\).  The augmented state obeys the same dynamics given by \(\ref{eq:nn}\), the only difference being that when we train the network on data from the real system, we do not include the hidden states \(h_i\) in the error function.  The hidden states are therefore free to change however they wish, as long as they result in the visible states \(x_i\) approximating the real dynamics.  After the model has been trained, one can perform Delayed DDP on this model.  One interesting point which arises is that since Delayed DDP is being applied to the augmented states, feedback gains will be obtained which correspond to both the hidden and visible states.  When applying these feedback gains to the real system, we can simply throw out the feedback gains which correspond to the hidden states.

\begin{figure}
\centering
\includegraphics[width = 0.48 \textwidth]{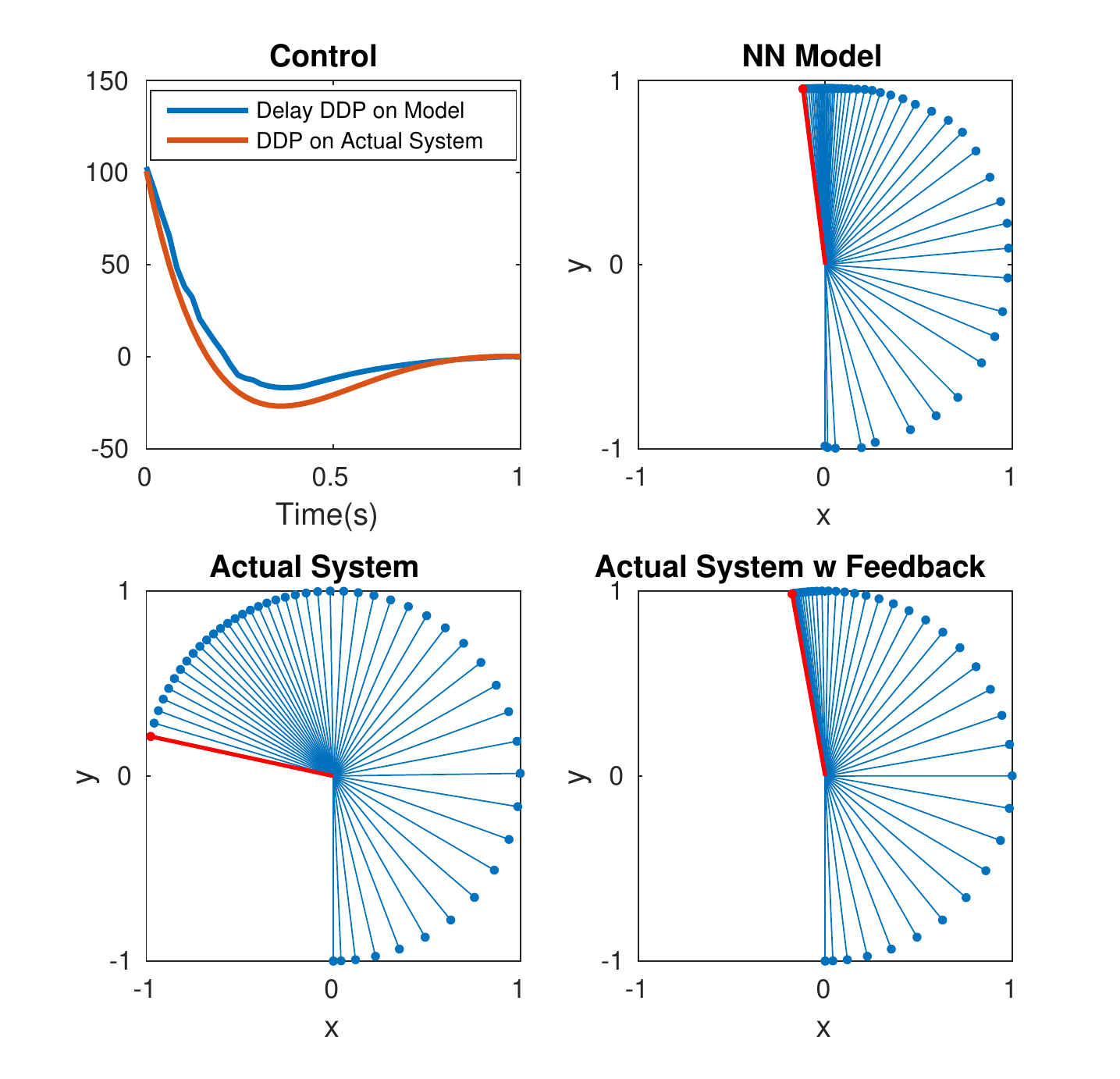}
\caption{Results demonstrating successful modeling and control of a neural network model for an inverted pendulum system where the model is trained on position information only.  Top Left:  Control sequence obtained from Delayed DPP using the neural network model, compared with control sequence obtained from classic DDP applied to the actual system.  Top Right: The behavior of the neural network model in response to the Delayed DDP control.  The red line shows the final position.  Bottom Left:  Behavior of the actual system when the Delayed DDP control obtained from the neural network model is applied, without using feedback gains.  Bottom Right:  Behavior of the actual system when the Delayed DDP control obtained from the neural network model is applied, utilizing the feedback gains corresponding to the current position only.}
\label{fig:pend}
\end{figure}
We trained a neural network with delays to model the dynamics of a pendulum from position information only, then use Delayed DDP to find a control which swings the pendulum into an inverted position.  The training dataset consisted of 12,500 1 second trajectories, simulated with a 20ms resolution, with each trajectory starting from the stable hanging position and perturbed by a random torque input.  This data consisted of the sequences of x and y coordinates of the pendulum bob scaled to the interval \([-1,1]\).  We did not include velocity information in the data, necessitating the use of a delayed system to infer velocity information.  The input was chosen to be a set of sinusoids with random frequency, phase, and amplitude, along with a set of uniformly distributed random control inputs.  A delay of \(k=3\) timesteps was used, so the input state consisted of 4 timesteps of data.  We used a neural network size of 32 units, 2 of which were the visible states.  Training was performed with the Adam method using a batch size of 128 and a total of 1000 epochs \cite{Kingma2014}.
  
Once the model was trained, Delayed DDP was applied to the neural network model.  The loss function used was the error in the angle of the pendulum, found by taking the inverse tangent of the x and y coordinates given by the visible states.  A small control cost was used to keep the control within the range of control inputs found in the training dataset.  Convergence occurred after about 15 iterations.  After finding the optimal control for the neural network model, this control was then applied to the real system, along with feedback gains.  Since the neural network model contains both delays and hidden states, we obtain extra feedback gains which are not usable on the real system.  We only use the feedback gains which correspond to the position at the current time.  The feedback gains scale the error between the states of the real system and the neural network model's states.  Figure \ref{fig:pend} shows the results comparing the control applied to the model, the real system, and the real system utilizing the feedback gains obtained from Delayed DDP.  Delayed DDP gives a control which successfully controls the model.  Comparing this control to the control obtained by classic DDP on the actual system shows that a similar solution has been found.  Applying the Delayed DDP control to the real system gives a less optimal result, due to slight modeling error in the neural network.  However, applying the feedback gains allows the successful control of the real system to the desired optimal trajectory, thereby compensating for the modeling error.  It is important to note here that the feedback gains applied to the real system are for position alone, since the neural network encodes position information only and does not encode the velocity of the pendulum.

\section{CONCLUSION}
We derived a differential dynamic programming algorithm for systems with delays in their state.  This allows us to leverage the power of DDP on a broad class of time-delayed systems, including neural network models which incorporate some past segment of history into their states.  We demonstrated the algorithm on a two-tank CSTR system, as well as a neural network trained to model an inverted pendulum from position information only, and a neural network trained to model an inverted cart-pole system.  We showed that leveraging the feedback gains that DDP gives allows us to create a control which is  more tolerant to noise and model error.

  Future research includes extension of differential dynamic programming for the case of nonlinear stochastic delayed systems. In addition, min-max and risk sensitive control formulation will also be under consideration. Finally, applications to real systems is ongoing work.




\bibliographystyle{IEEEtran}
\bibliography{delayddp_ieeetran_draft}

\begin{thebibliography}{10}
\providecommand{\url}[1]{#1}
\csname url@samestyle\endcsname
\providecommand{\newblock}{\relax}
\providecommand{\bibinfo}[2]{#2}
\providecommand{\BIBentrySTDinterwordspacing}{\spaceskip=0pt\relax}
\providecommand{\BIBentryALTinterwordstretchfactor}{4}
\providecommand{\BIBentryALTinterwordspacing}{\spaceskip=\fontdimen2\font plus
\BIBentryALTinterwordstretchfactor\fontdimen3\font minus
  \fontdimen4\font\relax}
\providecommand{\BIBforeignlanguage}[2]{{%
\expandafter\ifx\csname l@#1\endcsname\relax
\typeout{** WARNING: IEEEtran.bst: No hyphenation pattern has been}%
\typeout{** loaded for the language `#1'. Using the pattern for}%
\typeout{** the default language instead.}%
\else
\language=\csname l@#1\endcsname
\fi
#2}}
\providecommand{\BIBdecl}{\relax}
\BIBdecl

\bibitem{Lenz}
I.~Lenz, R.~Knepper, and A.~Saxena, ``{DeepMPC : Learning Deep Latent Features
  for Model Predictive Control},'' \emph{Robotics: Science and Systems}, 2015.

\bibitem{Wahlstrom2015a}
N.~Wahlstr{\"{o}}m, T.~B. Sch{\"{o}}n, and M.~P. Deisenroth, ``{From Pixels to
  Torques: Policy Learning with Deep Dynamical Models},'' \emph{arXiv preprint
  arXiv:1502.02251}, Feb 2015.

\bibitem{Belo}
F.~D. Marques, L.~D. F. R.~D. Souza, D.~C. Rebolho, a.~S. Caporali, E.~M. Belo,
  and R.~L. Ortolan, ``{Application of time-delay neural and recurrent neural
  networks for the identification of a hingeless helicopter blade flapping and
  torsion motions},'' \emph{Journal of the Brazilian Society of Mechanical
  Sciences and Engineering}, vol.~27, no.~2, pp. 97--103, 2005.

\bibitem{GL1961}
V.~G. Boltyanskiy, ``{The Maximum Principle in the Theory of Optimal
  Processes},'' \emph{Doklady Akademii Nauk}, no. 136, pp. 1--5, 1961.

\bibitem{Guinn1976}
T.~Guinn, ``{Reduction of delayed optimal control problems to nondelayed
  problems},'' \emph{Journal of Optimization Theory and Applications}, vol.~18,
  no.~3, pp. 371--377, Mar 1976.

\bibitem{Gollmann2009}
B.~Houska, H.~J. Ferreau, and M.~Diehl, ``{ACADO toolkit-An open-source
  framework for automatic control and dynamic optimization},'' \emph{Optimal
  Control Applications and Methods}, vol.~32, no.~3, pp. 298--312, Jul 2011.

\bibitem{Dadebo1992}
S.~Dadebo and R.~Luus, ``{Optimal control of time-delay systems by dynamic
  programming},'' \emph{Optimal Control Applications and Methods}, vol.~13,
  no.~1, pp. 29--41, Jan 1992.

\bibitem{Hwang1999}
C.~Hwang and J.~Lin, ``\BIBforeignlanguage{en}{{An improved computational
  scheme for solving dynamic optimization problems with iterative dynamic
  programming}},'' \emph{\BIBforeignlanguage{en}{Journal of the Chinese
  Institute of Engineers}}, vol.~22, no.~4, pp. 409--421, Jun 1999.

\bibitem{Mayne1966}
D.~Mayne, ``\BIBforeignlanguage{en}{{A Second-order Gradient Method for
  Determining Optimal Trajectories of Non-linear Discrete-time Systems}},''
  \emph{\BIBforeignlanguage{en}{International Journal of Control}}, vol.~3,
  no.~1, pp. 85--95, Jan 1966.

\bibitem{Tassa}
Y.~Tassa, T.~Erez, and E.~Todorov, ``\BIBforeignlanguage{English}{{Synthesis
  and stabilization of complex behaviors through online trajectory
  optimization}},'' in \emph{\BIBforeignlanguage{English}{IEEE International
  Conference on Intelligent Robots and Systems}}.\hskip 1em plus 0.5em minus
  0.4em\relax IEEE, Oct 2012, pp. 4906--4913.

\bibitem{Todorov2005}
E.~Todorov, ``{A generalized iterative LQG method for locally-optimal feedback
  control of constrained nonlinear stochastic systems},'' \emph{American
  Control Conference, 2005.}, pp. 300--306, 2005.

\bibitem{Tassa2014}
Y.~Tassa, N.~Mansard, and E.~Todorov, ``{Control-Limited Differential Dynamic
  Programming},'' \emph{International Conference on Robotics and Automation},
  pp. 1--8, 2014.

\bibitem{Theodorou2010}
E.~Theodorou, Y.~Tassa, and E.~Todorov, ``{Stochastic Differential Dynamic
  Programming},'' in \emph{Proceedings of the 2010 American Control
  Conference}, vol.~0, no.~1.\hskip 1em plus 0.5em minus 0.4em\relax IEEE, Jun
  2010, pp. 1125--1132.

\bibitem{Kobilarov}
M.~Kobilarov, D.~Ta, and F.~Dellaert, ``{Differential Dynamic Programming for
  Optimal Estimation},'' \emph{IEEE International Conference on Robotics and
  Automation}, 2015.

\bibitem{Pan2014a}
Y.~Pan and E.~Theodorou, ``{Probabilistic Differential Dynamic Programming},''
  in \emph{Advances in Neural Information Processing Systems}, 2014, pp.
  1907--1915.

\bibitem{Tassa2008a}
Y.~Tassa, T.~Erez, and W.~D. Smart, ``{Receding Horizon Differential Dynamic
  Programming},'' in \emph{Advances in Neural Information Processing Systems},
  2008, pp. 1465--1472.

\bibitem{Polycarpou1991}
M.~Polycarpou and P.~Ioannou, \emph{{Identification and control of nonlinear
  systems using neural network models: Design and stability analysis}}.\hskip
  1em plus 0.5em minus 0.4em\relax University of Southern Calif., 1991.

\bibitem{Funahashi1993}
K.~Funahashi and Y.~Nakamura, ``{Approximation of dynamical systems by
  continuous time recurrent neural networks},'' \emph{Neural Networks}, vol.~6,
  no.~6, pp. 801--806, 1993.

\bibitem{Sutskever2010}
I.~Sutskever and G.~Hinton, ``{Temporal-Kernel Recurrent Neural Networks},''
  \emph{Neural Networks}, vol.~23, no.~2, pp. 239--243, Mar 2010.

\bibitem{Schmidhuber2014}
J.~Schmidhuber, ``{Deep Learning in Neural Networks: An Overview},''
  \emph{Neural Networks}, vol.~61, pp. 85--117, Apr 2015.

\bibitem{Hartmann2012}
C.~Hartmann and J.~Boedecker, ``{Real-Time Inverse Dynamics Learning for
  Musculoskeletal Robots based on Echo State Gaussian Process Regression.}''
  \emph{Robotics: Science and Systems}, 2012.

\bibitem{Pascanu2013}
R.~Pascanu, C.~Gulcehre, K.~Cho, and Y.~Bengio, ``{How to Construct Deep
  Recurrent Neural Networks},'' \emph{arXiv preprint arXiv:1312.6026}, Dec
  2013.

\bibitem{Bengio2015a}
S.~Bengio, O.~Vinyals, N.~Jaitly, and N.~Shazeer, ``{Scheduled Sampling for
  Sequence Prediction with Recurrent Neural Networks},'' \emph{Advances in
  Neural Information Processing Systems}, pp. 1171--1179, Jun 2015.

\bibitem{Venkatraman2015}
A.~Venkatraman, M.~Hebert, and J.~A. Bagnell, ``{Improving Multi-step
  Prediction of Learned Time Series Models},'' \emph{Twenty-Ninth AAAI
  Conference on Artificial Intelligence}, 2015.

\bibitem{Kingma2014}
D.~P. Kingma and J.~L. Ba, ``{Adam: a Method for Stochastic Optimization},''
  \emph{International Conference on Learning Representations}, pp. 1--13, Dec
  2015.

\end{thebibliography}
%


\end{document}